\documentclass[prd,twocolumn,showpacs,amsmath,amssymb]{revtex4}
\usepackage{graphicx}
\usepackage{dcolumn}
\usepackage{bm}
\usepackage{rotating}
\usepackage{color}
\usepackage{subfigure}
\usepackage{pstricks}
\usepackage{pst-node}
\usepackage{epstopdf}
\usepackage{overpic}

\begin{document}
\normalsize
\parskip=5pt plus 1pt minus 1pt

\title{\boldmath Observation of the decay $\psi(3686)$ $\rightarrow$ $\Lambda\bar\Sigma^{\pm}\pi^{\mp}+c.c.$}

\author{\small
M.~Ablikim$^{1}$, M.~N.~Achasov$^{8,a}$, X.~C.~Ai$^{1}$, O.~Albayrak$^{4}$, D.~J.~Ambrose$^{41}$, F.~F.~An$^{1}$, Q.~An$^{42}$, J.~Z.~Bai$^{1}$, R.~Baldini Ferroli$^{19A}$, Y.~Ban$^{28}$
% ...
\\
\vspace{0.2cm}
(BESIII Collaboration)\\
\vspace{0.2cm} {\it
$^{1}$ Institute of High Energy Physics, Beijing 100049, People's Republic of China\\
% ...
$^{4}$ Carnegie Mellon University, Pittsburgh, Pennsylvania 15213, USA\\
% ...
$^{8}$ G.I. Budker Institute of Nuclear Physics SB RAS (BINP), Novosibirsk 630090, Russia\\
% ...
$^{19}$ (A)INFN Laboratori Nazionali di Frascati, I-00044, Frascati, Italy; (B)INFN and University of Perugia, I-06100, Perugia, Italy\\
% ...
$^{28}$ Peking University, Beijing 100871, People's Republic of China\\
% ...
$^{41}$ University of Rochester, Rochester, New York 14627, USA\\
$^{42}$ University of Science and Technology of China, Hefei 230026, People's Republic of China\\
% ...
$^{a}$ Also at the Novosibirsk State University, Novosibirsk, 630090, Russia\\
% ...
}
}

\begin{abstract}

Using a sample of $1.06\times10^{8}$ $\psi(3686)$ events collected with the BESIII detector,
we present the first observation of the decays of
$\psi(3686)$ $\rightarrow$ $\Lambda\bar\Sigma^{+}\pi^{-}+c.c.$ and
$\psi(3686)$ $\rightarrow$ $\Lambda\bar\Sigma^{-}\pi^{+}+c.c.$.
The branching fractions are measured to be $\mathcal{B}(\psi(3686) \rightarrow
\Lambda\bar\Sigma^{+}\pi^{-} + c.c.)=(1.40\pm 0.03 \pm 0.13 )\times10^{-4}$ and
$\mathcal{B}(\psi(3686) \rightarrow
\Lambda\bar\Sigma^{-}\pi^{+}+c.c.)=(1.54\pm 0.04 \pm 0.13 )\times10^{-4}$, where the first errors are statistical and the second ones systematic.

\end{abstract}


\maketitle

\section{Introduction}

Charmonium decays provide an ideal laboratory where our understanding
of nonperturbative Quantum Chromodynamics (QCD) and its interplay with
perturbative QCD can be tested~\cite{HPS}.
Perturbative QCD~\cite{exp-qcd1,exp-qcd2} predicts
that the partial widths for $J/\psi$ and $\psi(3686)$ decays into an
exclusive hadronic state $h$ are proportional to the squares of the
$c\bar{c}$ wave-function overlap at zero quark separation, which are
well determined from the leptonic widths. Since the strong coupling
constant, $\alpha_{s}$, is not very different at the $J/\psi$ and
$\psi(3686)$ masses, it is expected that the $J/\psi$ and $\psi(3686)$
branching fractions of any exclusive hadronic state $h$ are related by
\begin{eqnarray}
Q_{h}=\frac{\mathcal{B}(\psi(3686)\rightarrow h)}{\mathcal{B}(J/\psi
\rightarrow h)} \cong \frac{\mathcal{B}(\psi(3686)\rightarrow
e^{+}e^{-})} {\mathcal{B}(J/\psi \rightarrow e^{+}e^{-})} \cong
12\%.\nonumber
\end{eqnarray}
This relation defines the "$12\%$ rule", which works reasonably well
for many specific decay modes. A large violation of this rule was
observed by later experiments~\cite{mark2-rhopi,bes2-2005,cleo-2005},
particularly in $\rho\pi$ decay. Recent
reviews~\cite{theory-expected1,theory-expected3}
of relevant theories and experiments conclude that current
theoretical explanations are unsatisfactory. Clearly, more
experimental results are desirable.

The study of baryon spectroscopy plays an important role in the
development of the quark model and in the understanding of
QCD~\cite{PDG}-\cite{MISS}. However, our knowledge on baryon spectroscopy is
limited; in particular the number of observed baryons is significantly
smaller than what is expected from the quark model. For a recent
review of baryon spectroscopy, see Ref.~\cite{MPR}.

Three body charmonium decays of $J/\psi$ and $\psi(3686)$ decays,
provide a complementary approach to study the internal structure of
light baryons with respect to the traditional pion (kaon) scattering
experiments.  Using 58 million $J/\psi$ events, the BESII
Collaboration reported the observation of a new N*
resonance~\cite{PNPI}, denoted as N(2065), in $J/\psi\rightarrow p
\bar{n}\pi^-+c.c.$, which was subsequently confirmed in
$J/\psi\rightarrow p\bar{p}\pi^0$~\cite{PPPI}. More recently, with 106
million $\psi(3686)$ events, two new structures, N(2300) and N(2570),
were observed at the BESIII experiment in $\psi(3686)\rightarrow
p\bar{p}\pi^0$ decay~\cite{PPPILYT}~\cite{CLEO}.  Not only excited
nucleons, but also baryons with one strange quark (eg. $\Lambda^*$ and
$\Sigma^*$) can be studied in $J/\psi$ and $\psi(3686)$ decays.

In this paper, we study $\psi(3686)$ $\rightarrow$
$\Lambda\bar\Sigma^{+}\pi^{-}+c.c.  $ and $\psi(3686)$ $\rightarrow$
$\Lambda\bar\Sigma^{-}\pi^{+}+c.c.  $, and measure the corresponding
branching fractions for the first time using $1.06\times10^{8}$
$\psi(3686)$ events collected with the Beijing Spectrometer (BESIII)
detector. Further, the branching fraction of $\psi(3686)
\rightarrow\Lambda\bar\Sigma^-\pi^+$ and that from $J/\psi$ decay are
used to test the ``$12\%$ rule''~\cite{exp-qcd1,exp-qcd2}.  Peaks are
observed around $1.5$ GeV$/c^{2}$ to $1.7$ GeV$/c^{2}$ in the
$\bar\Sigma^+\pi^-$ and
$\Lambda\pi^-$ mass spectra, which are indicative of $\Lambda^{*}$ and
$\Sigma^{*}$ states, respectively.

\section{DETECTOR AND MONTE CARLO SIMULATION}
The Beijing Electron Positron Collider (BEPCII)~\cite{BESIII} is a
double-ring e$^{+}$e$^{-}$ collider designed to provide a peak
luminosity of $10^{33}$ cm$^{-2}s^{-1}$ at a center of mass energy of
$3.77$~GeV.  The BESIII~\cite{BESIII} detector has a geometrical
acceptance of $93\%$ of $4\pi$ and has four main components: (1) A
small-cell, helium-based ($40\%$ He, $60\%$ C$_{3}$H$_{8}$) main drift
chamber (MDC) with $43$ layers providing an average single-hit
resolution of $135$~$\mu$m, and charged-particle momentum resolution
in a $1$~T magnetic field of $0.5\%$ at 1~GeV$/c$. (2) An
electromagnetic calorimeter (EMC) consisting of $6240$ CsI(Tl)
crystals in a cylindrical structure (barrel) and two endcaps. For
$1$~GeV photons, the energy resolution is $2.5\%$ ($5\%$) and the
position resolution is $6$~mm ($9$~mm) in the barrel (endcaps). (3) A
time-of-flight system (TOF) consisting of $5$-cm-thick plastic
scintillators, with $176$ detectors of $2.4$~m length in two layers in
the barrel and $96$ fan-shaped detectors in the endcaps. The barrel
(endcaps) time resolution of $80$~ps ($110$~ps) provides $2\sigma$
$K/\pi$ separation for momenta up to $\sim 1$~GeV$/c$. (4) The muon
system consisting of $1000$~m$^{2}$ of resistive plate chambers in $9$
barrel and $8$ endcap layers and providing a position resolution of
$2$~cm.

The optimization of the event selection and the estimation of
backgrounds are performed through Monte Carlo (MC) simulations. The
{\sc Geant4}~\cite{GEANT} based simulation software {\sc
Boost}~\cite{BOOST} includes the geometry and material description
of the BESIII spectrometer and the detector response and digitization
models, as well as the tracking of the detector running conditions
and performance. The production of the $\psi(3686)$ resonance is
simulated by the MC event generator {\sc kkmc}~\cite{KKMC1, KKMC2},
while the decays are generated by {\sc EvtGen}~\cite{EvtGen} for
known decay modes with branching fractions being set to world
average values~\cite{PDG}, and by {\sc LundCharm}~\cite{LUNDCHARM}
for the remaining unknown decays.

\section{Event selection}

In this analysis, the charge-conjugate reaction is always implied
unless explicitly mentioned. The $\bar\Sigma^-$ is reconstructed in its $\bar{p}
\pi^{0}$ and $\bar{n}\pi^{-}$ decay modes, and $\bar\Sigma^{+}$, $\Lambda$ and
$\pi^0$ are reconstructed in $\bar\Sigma^{+}\rightarrow \bar{n}\pi^{+}$,
$\Lambda\rightarrow p\pi^-$ and $\pi^0\rightarrow \gamma\gamma$. The
possible final states of $\psi(3686)\rightarrow {\Lambda}
\bar{\Sigma}^+\pi^-$ and $\psi(3686)\rightarrow{\Lambda}
\bar{\Sigma}^-\pi^+$ are then $p\pi^{-}\pi^{-}\pi^{+}\bar n$ and
$\gamma\gamma p\bar p\pi^{-}\pi^{+}$.  The following common selection
criteria, including charged track selection, particle identification
and $\Lambda$ reconstruction, are used to select candidate events.

Candidate events must have four charged tracks with zero net
charge. Tracks, reconstructed from the MDC hits, must have a polar
angle $\theta$ in the range $|\cos\theta|<0.93$ and pass within 20 cm
of the interaction point in the beam direction and within 10 cm in the
plane perpendicular to the beam. The pion produced directly from
$\psi(3686)$ decays must have its point of closest approach to the
beam line within 20 cm of the interaction point along the beam
direction and within $2.0$ cm in the plane perpendicular to the
beam. In order to suppress background events from
$\psi(3686)\rightarrow K^{0}_{S}\bar n\Lambda$, the point of closest
approach in the plane perpendicular to the beam is required to be
within $0.5$ cm in the cases of
$\bar\Sigma^-\rightarrow\bar{n}\pi^-+c.c.$ and
$\bar\Sigma^+\rightarrow\bar{n}\pi^++c.c.$.

For each charged track, both TOF and $dE/dx$ information are combined
to form Particle IDentification (PID) confidence levels for the $\pi$,
$K$, and $p$ hypotheses ($Prob(i)$, $i=\pi,K,p$). A charged track is
identified as a pion or proton if its $Prob$ is larger than those for
any other assignment. For all four channels with a neutron (or
anti-neutron), only one charged track is required to be identified as
a proton or anti-proton, and the other charged tracks are assigned as
pions.  In order to suppress background events from
$\psi(3686)\rightarrow \pi^{0} \pi^{0}J/\psi$ with $J/\psi \rightarrow
\Lambda\bar\Lambda$, the candidate pion should not be identified as an
anti-proton in the case of $\bar\Sigma^-\rightarrow\bar{n}\pi^-+c.c.$.
For $\bar{\Sigma}^- \rightarrow \bar{p} \pi^{0}+c.c.$, at least one of
the charged tracks should be identified as a proton or an anti-proton.

To reconstruct the decay $\Lambda\rightarrow p\pi^-$, a
vertex fitting algorithm is applied to all combinations of $p\pi^{-}$
pairs. If more than one $p\pi^{-}$ combination satisfies the
vertex fitting requirement, the pair with the mass closest
to $M(\Lambda)$ is chosen, where $M(\Lambda)$ is the nominal mass of
$\Lambda$~\cite{PDG}.


\subsection{$\psi(3686)\rightarrow {\Lambda}\bar\Sigma^-\pi^+\rightarrow p\bar{p}\pi^+\pi^-\gamma\gamma$}

Events selected with the above selection criteria and at least two
photon candidates are kept for further analysis. Photon
candidates, reconstructed by clustering EMC crystal energies, must
have a minimum energy of $25$ MeV for the barrel ($|\cos
\theta|<0.80$) and $50$ MeV for the endcap ($0.86< |\cos
\theta|<0.92$), must satisfy EMC cluster timing requirements to suppress
electronic noise and energy deposits unrelated to the event, and be
separated by at least $10^{\circ}$ from the nearest charged track
($20^{\circ}$ if the charged track is identified as an anti-proton) to
exclude energy deposits from charged particles.

\begin{figure*}[htbp]
\begin{center}
 \includegraphics[width=7.cm,height=5cm]{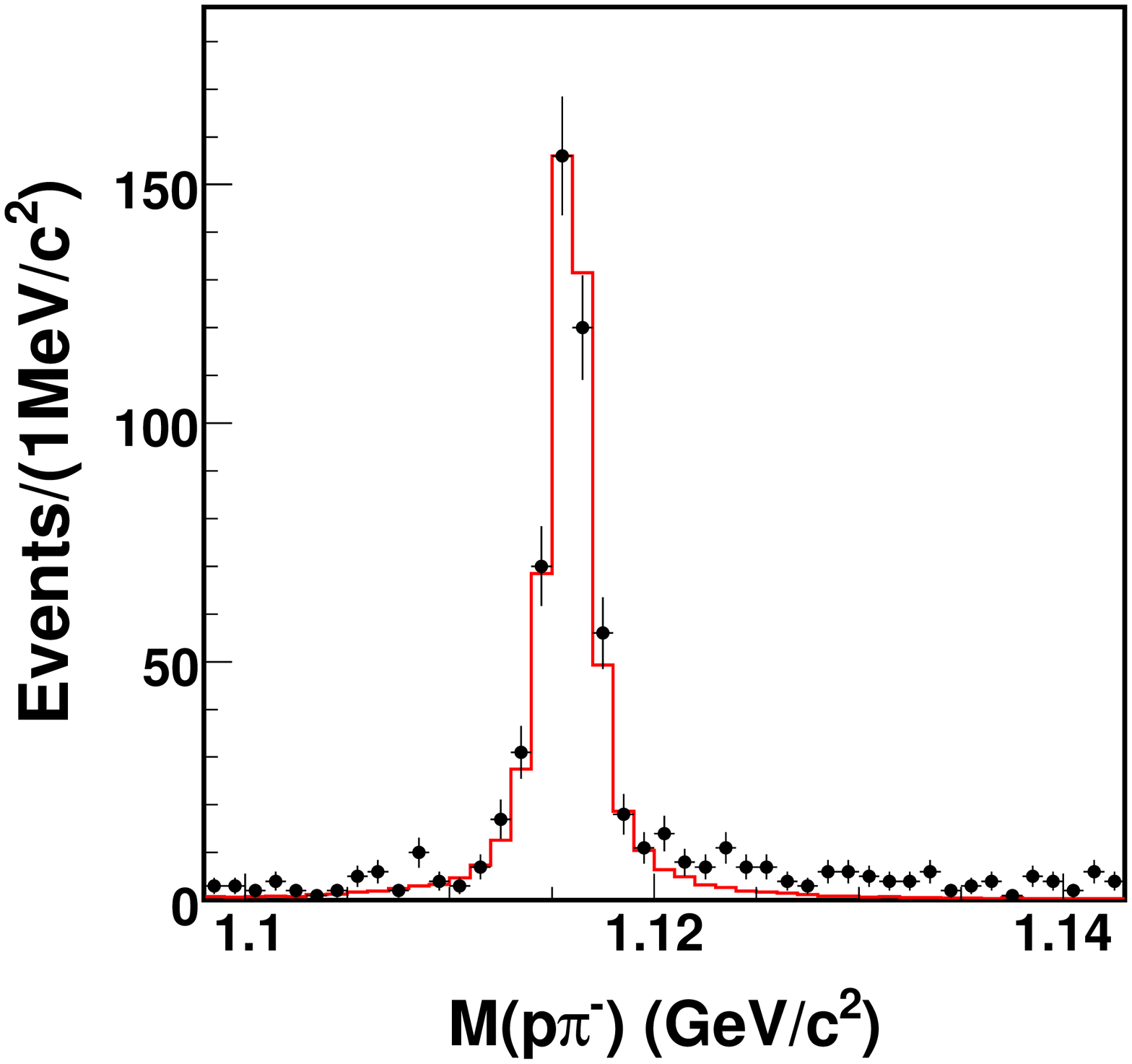}\put(-30,113){(a)}
 \includegraphics[width=7.cm,height=5cm]{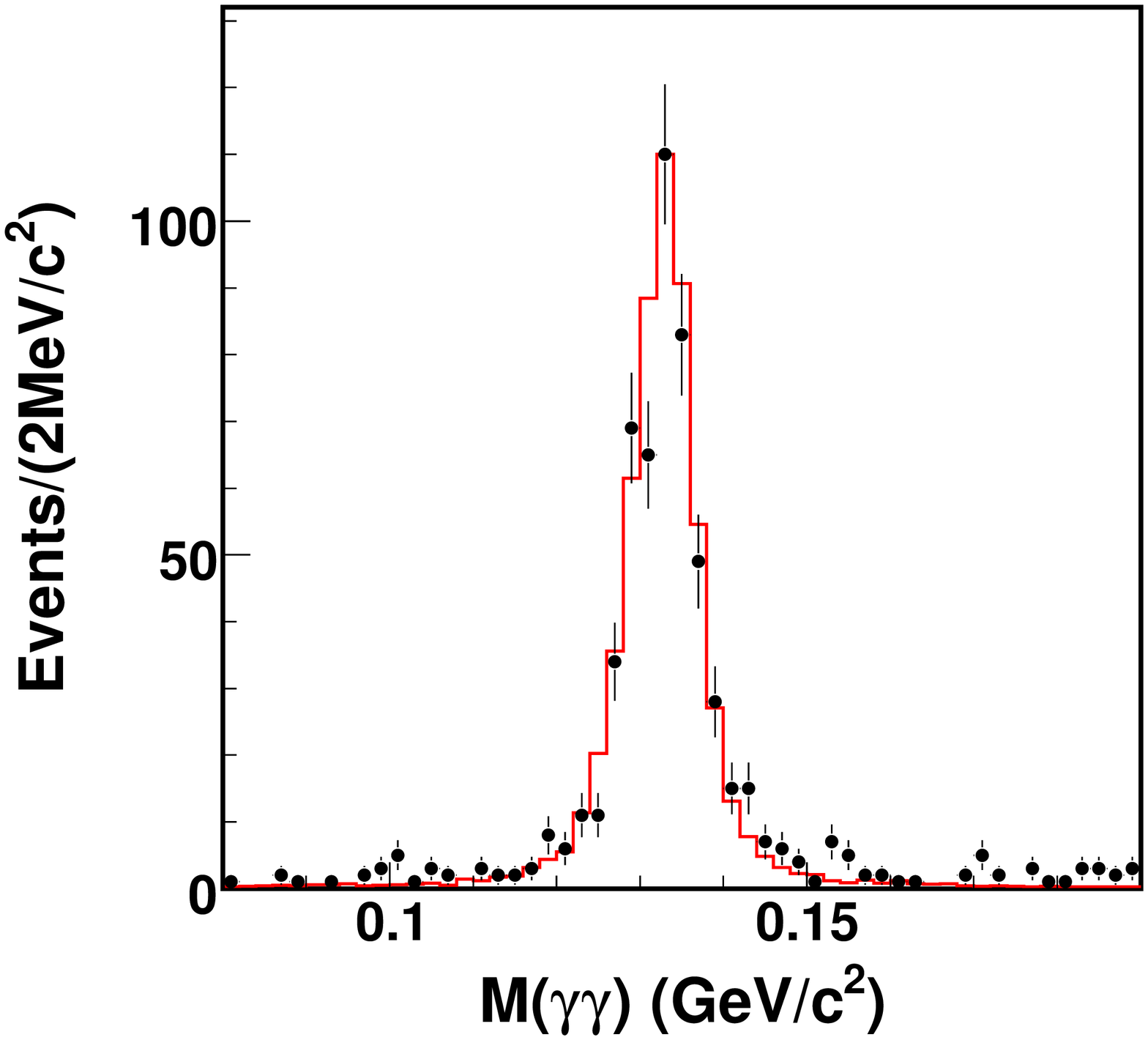}\put(-30,113){(b)}
\caption{The distributions of (a) $M(p\pi^-)$ and (b)
$M(\gamma\gamma)$. The crosses with error bars are data, and the
histograms are signal MC simulations without background
included.}
\label{1_phsp_Lambda}
\label{3_phsp_Pi0}
\end{center}
\end{figure*}

Figure~\ref{1_phsp_Lambda}(a) shows the $p\pi^-$ mass, $M(p\pi^-)$,
distribution for events that satisfy the $\Lambda$ vertex finding
algorithm. A clear peak at the $\Lambda$ mass is observed, and a
$\Lambda$ mass window requirement, $1.111$ GeV$/c^{2}<M(p\pi^-)<1.121$
GeV$/c^{2}$, is applied to extract the $\Lambda$ signal.


A four-constraint kinematic fit imposing momentum and energy
conservation is performed under the $\gamma\gamma p\bar
p\pi^{-}\pi^{+}$ hypothesis, and the chisquare ($\chi^2_{\gamma\gamma
p\bar p\pi^{-}\pi^{+}}$) is required to be less than 100. 
For events with more than two photons, all combinations are tried, and
the combination with the smallest $\chi^2_{\gamma\gamma p\bar
p\pi^{-}\pi^{+}}$ is retained. The $\pi^0$ is clearly seen in the
$\gamma\gamma$ mass, $M(\gamma\gamma)$, spectrum shown in
Fig.~\ref{3_phsp_Pi0}(b). The $\bar p\pi^0$ invariant mass spectrum
for events in the $\pi^0$ mass window ($0.12$
GeV$/c^{2}<M(\gamma\gamma)<0.145$ GeV$/c^{2}$) is shown in
Fig.~\ref{3_Fit_S}(a), where the $\bar\Sigma^-$ peak is seen.

To extract the number of $\bar\Sigma^-$ events, an unbinned maximum
likelihood fit is applied to the $\bar p\pi^0$ mass spectrum with a
double Gaussian function for the signal plus a second order Chebychev
polynomial as the background function. The fit, shown as the solid
line in Fig.~\ref{3_Fit_S}(a), yields $458\pm23$ $\bar{\Sigma}^-$
events, while the fit to the ${p}\pi^0$ mass distribution gives
$554\pm26$ $\Sigma^+$ events, as shown in Fig.~\ref{5_Fit_S}(b). The
non-peaking background can be well described by the events from
$\Lambda$ sideband. Fits of the $\Lambda$ and $\bar\Lambda$ sideband
events yield $18\pm5$ $\bar\Sigma^-$ and $13\pm5$ $\Sigma^+$ events.

\subsection{$\psi(3686)\rightarrow
{\Lambda} \bar\Sigma^+\pi^-(\Lambda\bar\Sigma^-\pi^+)\rightarrow p\bar{n}\pi^+\pi^-\pi^-$}

Neutrons cannot be fully reconstructed with the EMC information.
However, the distribution of mass recoiling against $p\pi^+\pi^-\pi^-$
tracks, $R(p\pi^+\pi^-\pi^-)$, for events with the recoiling mass and
the $\pi^+$ mass, $M(R(p\pi^+\pi^-\pi^-)\pi^+)$, inside the
$\bar{\Sigma}^+$ mass region ($1.186 < M(R(p\pi^+\pi^-\pi^-)\pi^+) <
1.208$ GeV/$c^2$), shown in Fig.~\ref{1_phsp_Miss}, has a significant
anti-neutron peak.  After requiring $|R(p\pi^+\pi^-\pi^-)-M(\bar
n)|<0.04$ GeV$/c^{2}$ $(3\sigma)$, where $M(\bar n)$ is the neutron
mass, a one-constraint kinematic fit with the recoil mass constrained
to the neutron mass is performed to improve the mass resolution, and
the chisquare $\chi^{2}(p\pi^{-}\pi^{-}\pi^{+}\bar n)$ is required to
be less than 20.

Using the same method described in Section $A$, we perform fits to the
$\bar{n}\pi^+$, ${n}\pi^-$, $n\pi^+$, and $\bar{n}\pi^-$ mass
distributions ($M(\bar{n}\pi^+)$, $M({n}\pi^-)$, $M({n}\pi^+)$ and
$M(\bar{n}\pi^-)$) to extract the number of $\bar\Sigma^+$,
${\Sigma}^-$, $\Sigma^+$ and $\bar{\Sigma}^-$ events and background
events from the $\Lambda$ sideband. Here, the $n$ and $\bar{n}$
momenta from the one-constraint kinematic fits above are used to
determine $M(\bar{n}\pi^+)$, $M({n}\pi^-)$, $M({n}\pi^+)$ and
$M(\bar{n}\pi^-)$.  The fits are shown in Figs.~\ref{1_Fit_S}(a) to
~\ref{6_Fit_S}(d), and the fit results are summarized in
Table~\ref{12_listallchannels}.

\section{Background study}

In this analysis, 106 million inclusive $\psi(3686)$ MC events are
used to investigate possible backgrounds from $\psi(3686)$ decays. The
results indicate that the background events mainly have an
approximately flat distribution. Since the background contributions to
the $\Sigma$ peak are not very significant, and the branching
fractions of some possible decay channels are not yet well measured,
background contributions are estimated from $\Lambda$ sidebands,
defined as $1.1027$ GeV/$c^2$ $<M(p\pi^{-})<1.1077$ GeV/$c^2$ and
$1.1237$ GeV/$c^2$ $<M(p\pi^{-})<1.1337$ GeV/$c^2$, and shown in
Fig.~\ref{1_scatter_LS}(a), where $M(p\pi^{-})$ is the $p\pi^{-}$
invariant mass. Fitting the $\Lambda$ sideband events in the same way
as the signal events, we obtain the numbers of background events,
summarized in Table~\ref{12_listallchannels}, which will be subtracted
in the calculation of the branching fractions.

To estimate the number of background events coming directly from the
$e^+e^-$ annihilation, the same analysis is performed on data taken
at center-of-mass energy of 3.65 GeV, where the number of background
events are also extracted by fitting the $\bar n \pi^{+}$ (or
$\bar{p}\pi^0$) mass spectrum. The background events are then
normalized to the $\psi(3686)$ data after taking into account the
luminosities and energy-dependent cross section of the quantum
electrodynamics (QED) processes,

\begin{equation}
 N_{QED}=\frac{\mathcal{L}_{3.686}}{\mathcal{L}_{3.650}}\times\frac{3.65^{2}}{3.686^{2}}\times N_{3.65}^{fit},
\end{equation}
 where $N_{QED}$ is the number of background events from QED
 processes, $\mathcal{L}_{3.686}=165 $ pb$^{-1}$ and
 $\mathcal{L}_{3.650}=44 $ pb$^{-1}$ are the integrated luminosities
 for $\psi(3686)$ data~\cite{LUM} and $3.65$ GeV data~\cite{NUMBER},
 and $N_{3.65}^{fit}$ is the number of selected events from continuum
 data.

\begin{figure*}[htbp]
      \includegraphics[width=7.cm,height=5cm]{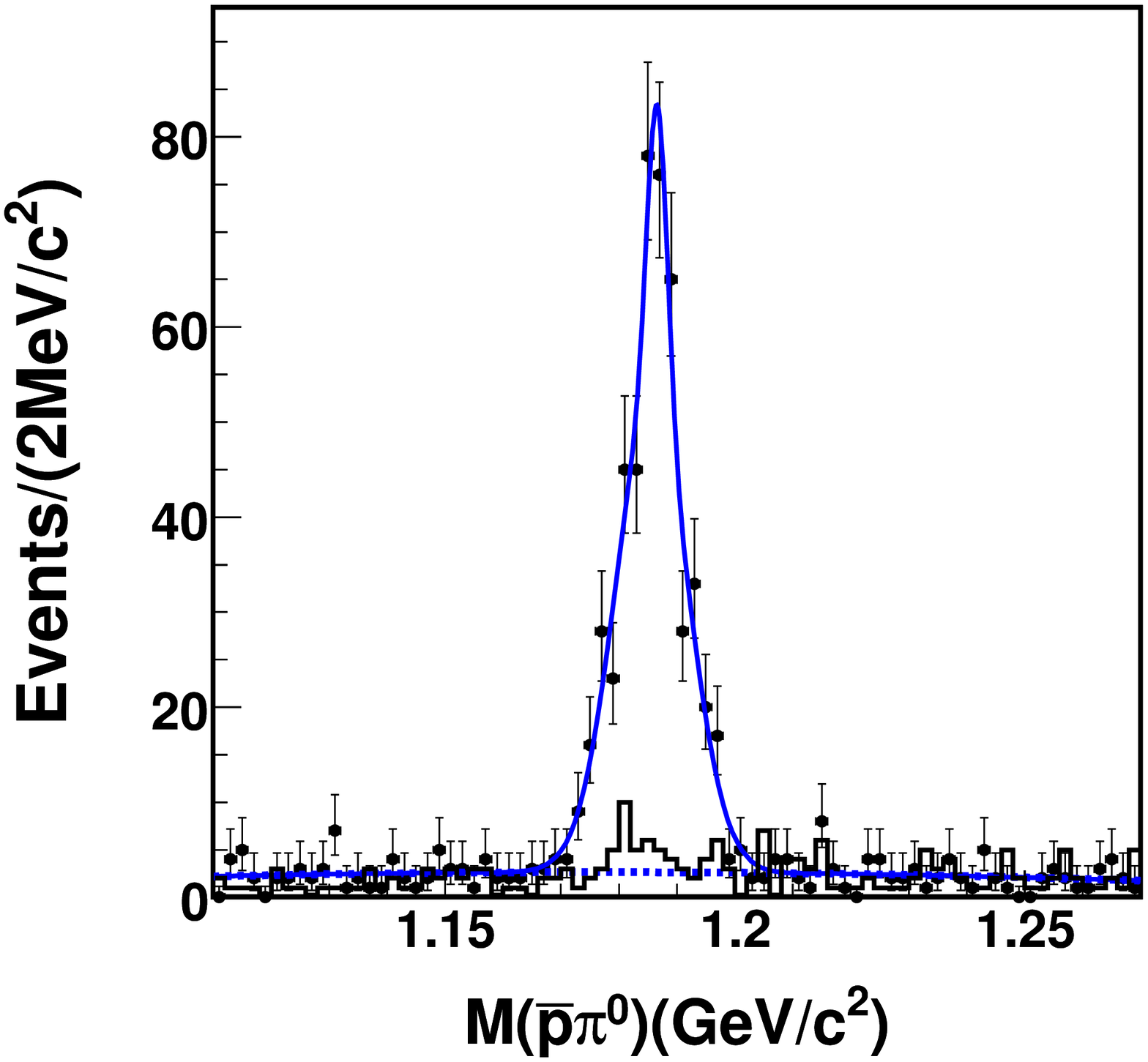}\put(-30,113){(a)}
      \label{3_Fit_S}
      \includegraphics[width=7.cm,height=5cm]{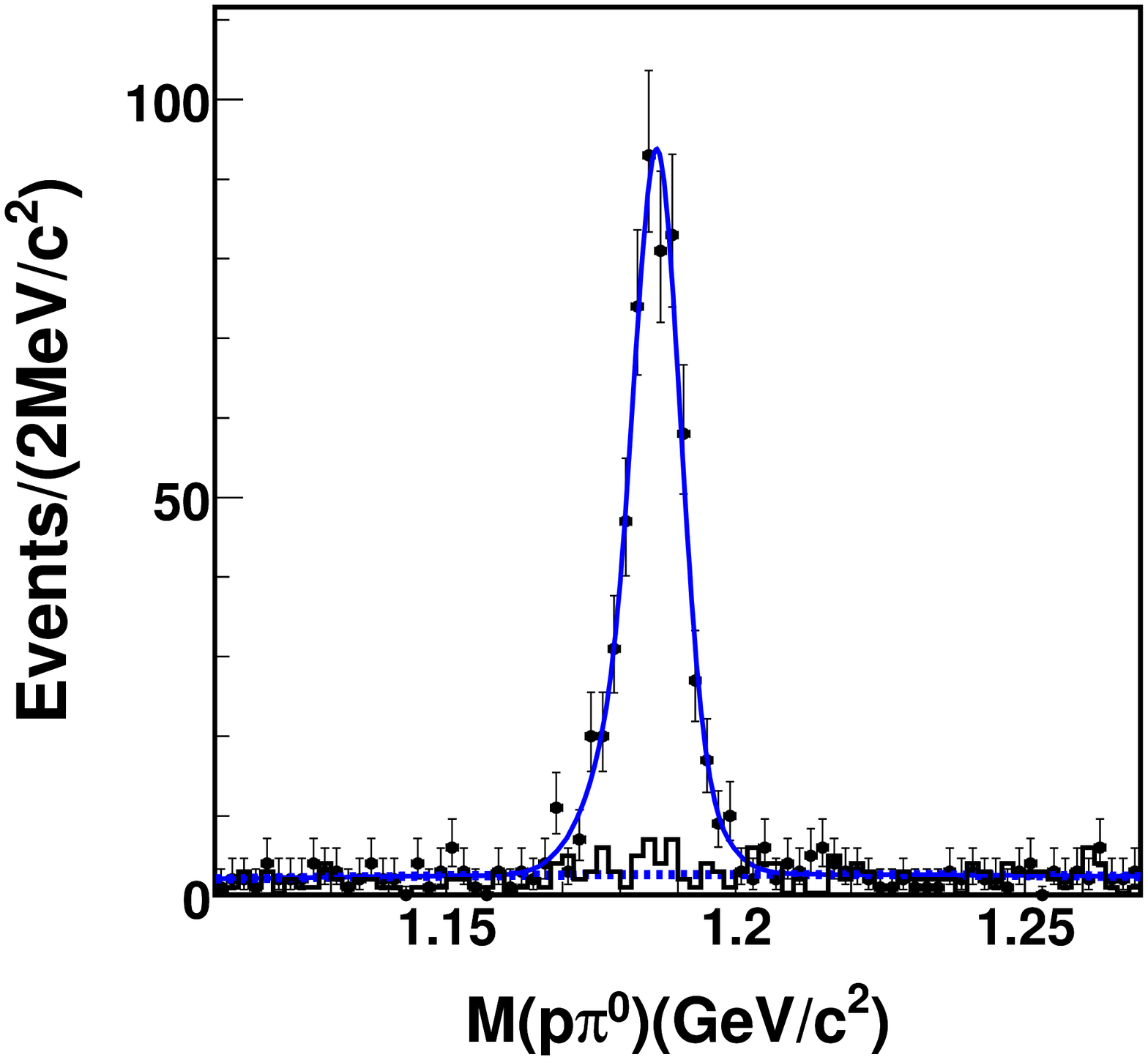}\put(-30,113){(b)}
      \label{5_Fit_S}
\caption{The distributions of (a) $M(\bar{p}\pi^0)$ and (b)
$M(p\pi^0)$. The crosses with error bars are data, the histograms
are background estimated with $\Lambda$($\bar{\Lambda}$)
sidebands, the solid lines are the fits described in the text, and the
dashed lines are the fits of background.}
\label{3_Fit_S}
\label{5_Fit_S}
\end{figure*}

\begin{figure*}[htbp]
      \includegraphics[width=7.cm,height=5cm]{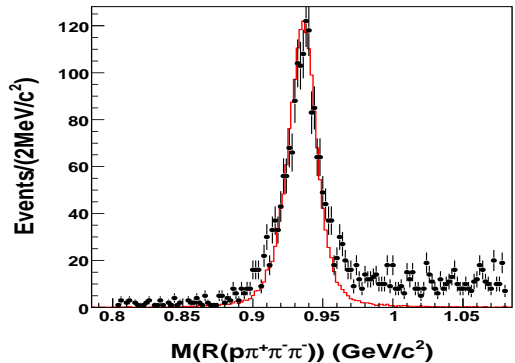}
      \label{1_phsp_Miss}
\caption{The distribution of the mass recoiling against
$p\pi^{+}\pi^{-}\pi^{-}$, where the crosses with error bars are
data and the histogram the MC simulation of signal events.}
\label{1_phsp_Miss}
\end{figure*}

\begin{figure*}[htbp]
      \includegraphics[width=7cm,height=5cm]{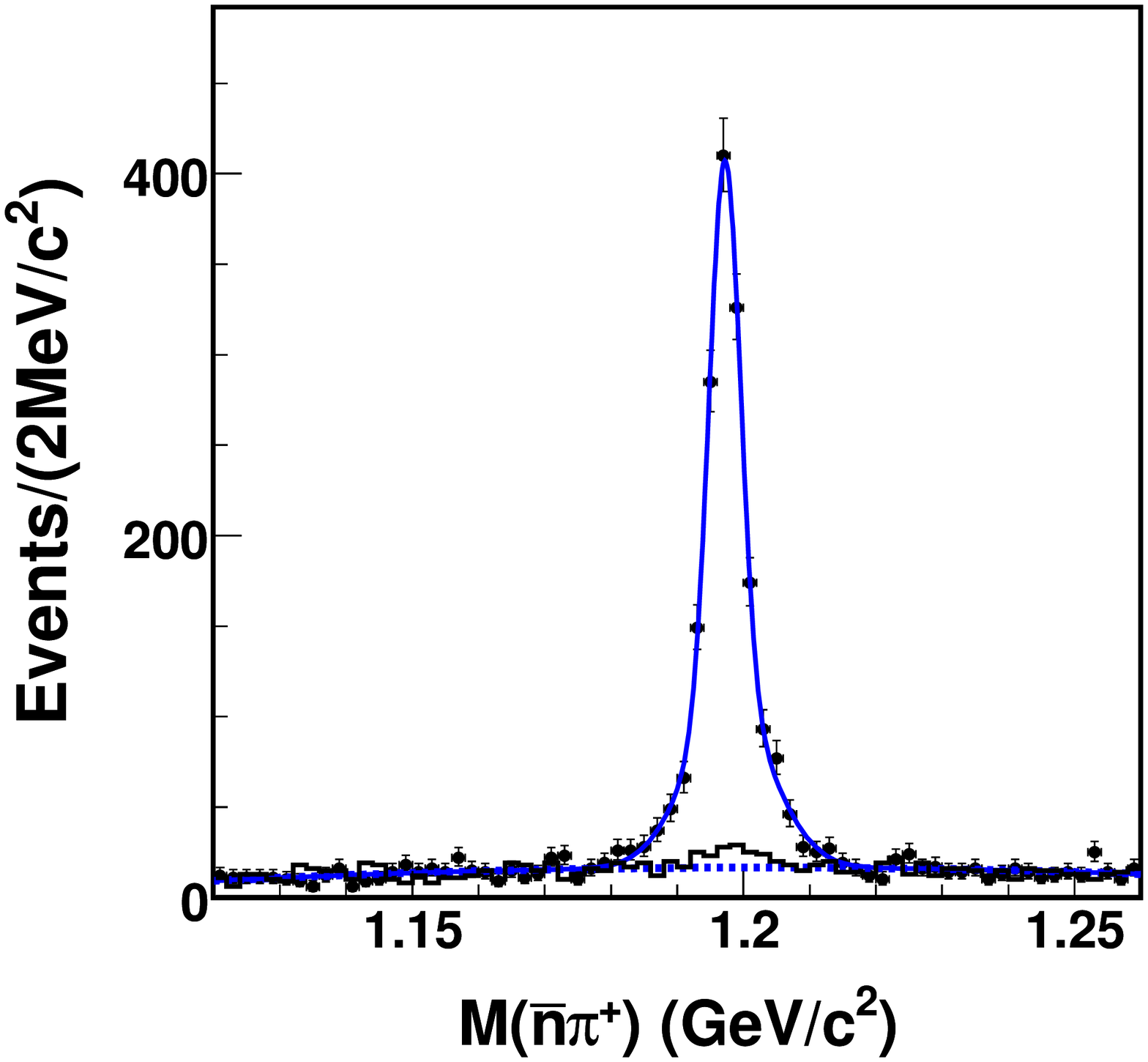}\put(-30,113){(a)}
      \label{1_Fit_S}
      \includegraphics[width=7cm,height=5cm]{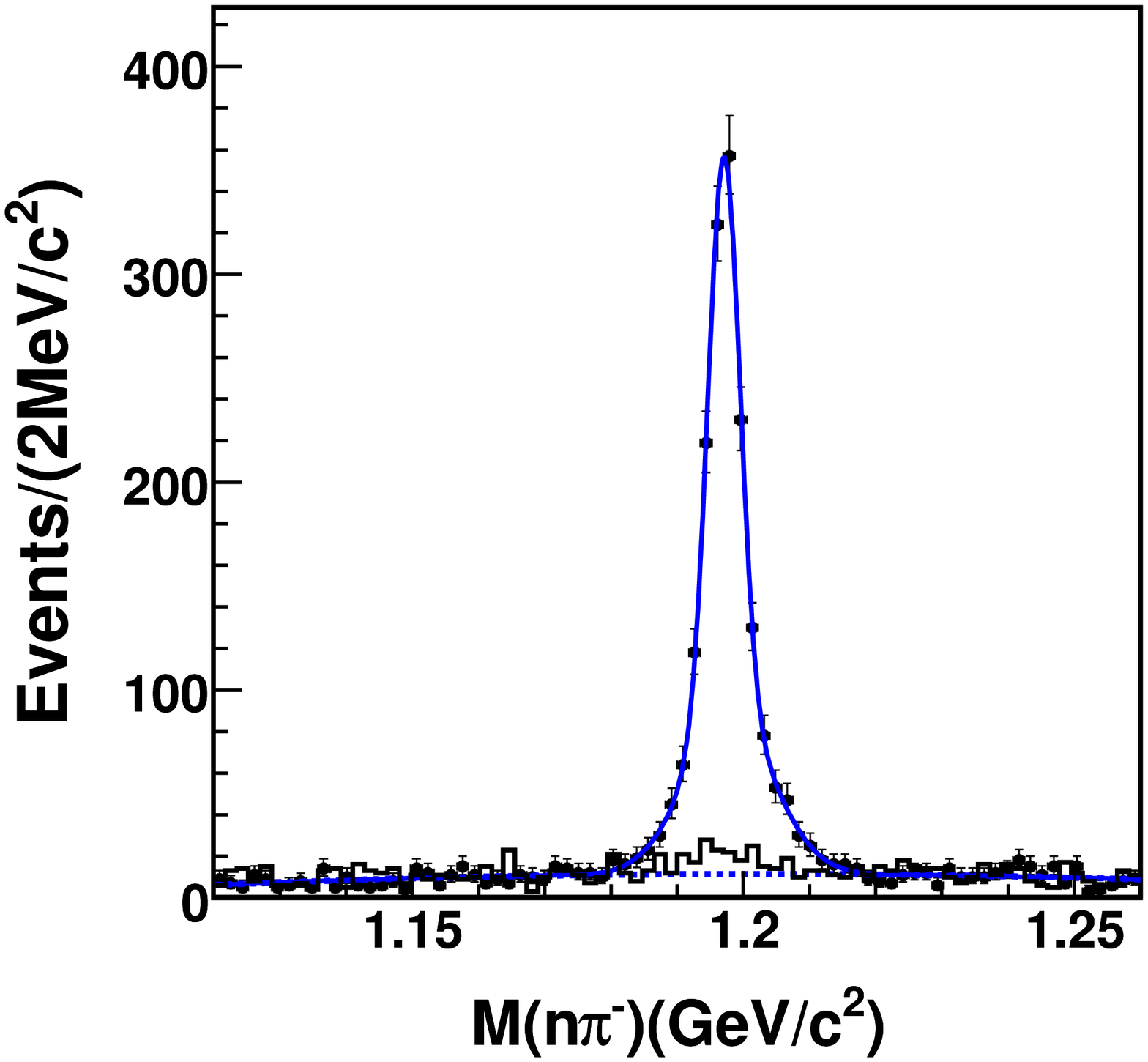}\put(-30,113){(b)}\\
      \label{6_Fit_S}
      \includegraphics[width=7cm,height=5cm]{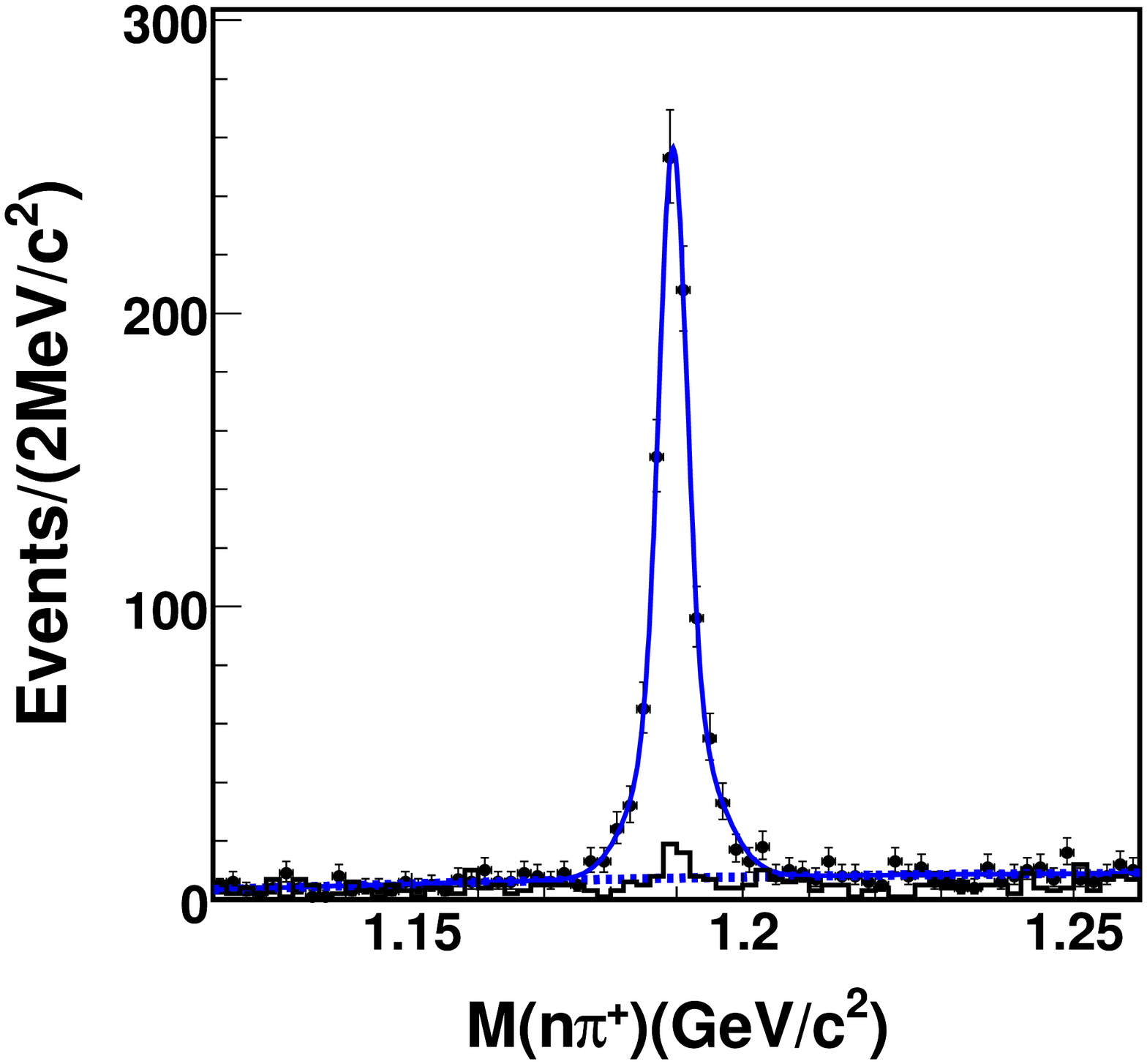}\put(-30,113){(c)}
      \label{4_Fit_S}
      \includegraphics[width=7cm,height=5cm]{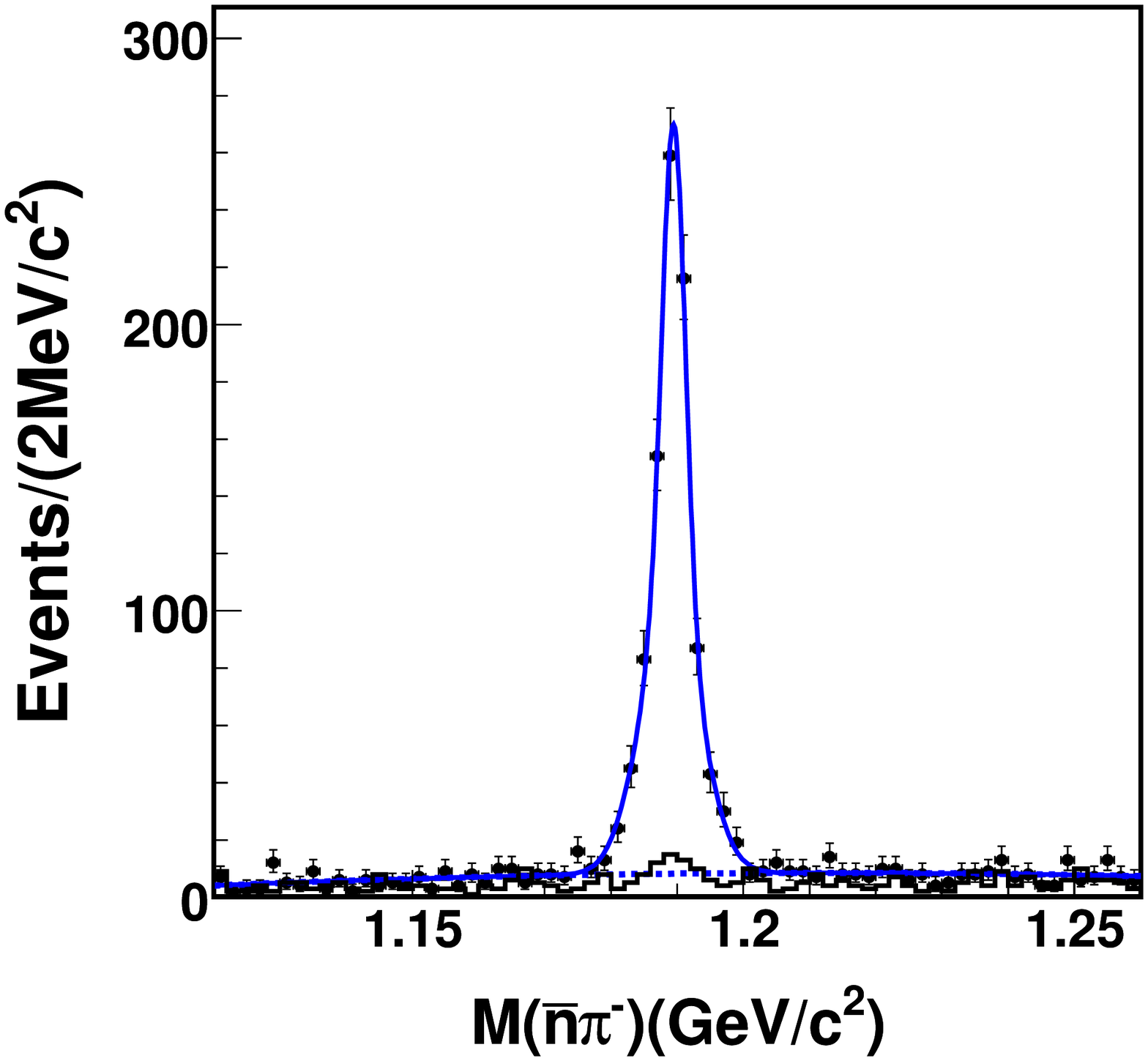}\put(-30,113){(d)}\\
      \label{2_Fit_S}

\caption{ The distributions of (a)
$M(\bar{n}\pi^+)$, (b) $M({n}\pi^-)$, (c) $M(n\pi^+)$, and
(d)$M(\bar{n}\pi^-)$. The crosses with error bars are data, the
histograms are background estimated with $\Lambda$($\bar{\Lambda}$)
sidebands, the solid lines are the fits described in the text, and the
dashed lines are the fits of background.}
\label{1_Fit_S}
\label{2_Fit_S}
\label{4_Fit_S}
\label{6_Fit_S}
\end{figure*}


\section{Detection efficiency determination}

To determine the detection efficiencies, possible intermediate states
decaying into $\bar\Sigma\pi$ and $\Lambda\pi$ are
investigated. Figure~\ref{1_pwafit_dalitzData}(b) is the Dalitz plot
of selected $\psi(3686)\rightarrow {\Lambda}
\bar\Sigma^+\pi^-\rightarrow p\bar{n}\pi^+\pi^-\pi^-$ candidates,
where clear clusters indicate that this process is mediated by excited
baryons. The two dimensional ${\Lambda}-\bar\Sigma$ sidebands, shown
as the boxes in Fig.~\ref{1_scatter_LS}(a), are used to estimate the
number of background events, and the background distributions, shown
as shaded histograms in Figs.~\ref{1_pwafit_LPi}(a),
\ref{1_pwafit_SPi}(b) and \ref{1_pwafit_LS}(c), indicate that the
structures are not from background events. The $\Lambda \pi$ and
$\Sigma \pi$ invariant mass spectra, shown in
Fig.~\ref{1_pwafit_LPi}(a) and Fig.~\ref{1_pwafit_SPi}(b), indicate
$\Lambda^*$ and $\Sigma^*$ structures, eg. peaks around $1.4$
GeV$/c^{2}$ to $1.7$ GeV$/c^{2}$ in the invariant mass distributions
of $\Lambda\pi^-$ and $\bar\Sigma^+\pi^-$, that clearly deviate from
what is expected according to phase space. In order to determine the
correct detection efficiency, a Partial Wave Analysis (PWA) is
performed based on an unbinned maximum likelihood fit~\cite{PNPI}. As shown in Fig.~\ref{1_pwafit_LPi}, the background contamination is small and is ignored in the PWA. Sixteen possible intermediate excited states
($\Lambda(1810)$, $\Lambda(1800)$, $\Lambda(1670)$, $\Lambda(1600)$,
$\Lambda(1405)$, $\Lambda(1116)$, $\Lambda(2325)$, $\Lambda(1890)$,
$\Lambda(1690)$, $\Lambda(1520)$, $\Lambda(1830)$, $\Lambda(1820)$,
$\Sigma(1660)$, $\Sigma(1670)$, $\Sigma(1580)$ and $\Sigma(1385)$)
with at least two stars according to the PDG ~\cite{PDG} are
included in the PWA. In the global fit, all of these resonances are described
with Breit-Wigner functions, and the masses and widths are fixed to
the world average~\cite{PDG}. A comparison of the data and global
fitting results, shown in Fig.~\ref{1_pwafit_LPi}, indicates that the
PWA results are consistent with data. A similar PWA is also performed
for the decays $\psi(3686)\rightarrow
{\Lambda}\bar\Sigma^-\pi^+\rightarrow p\bar{p}\pi^+\pi^-\gamma\gamma$,
and the results are also in agreement with data. Finally the MC
samples of $\psi(3686)\rightarrow {\Lambda}\bar\Sigma^+\pi^-$ and
$\psi(3686)\rightarrow {\Lambda}\bar\Sigma^-\pi^+$ are generated
according to the PWA results, and the detection efficiencies are
determined by fitting the $\Sigma$ signal and $\Lambda$ sideband events and presented in Table~\ref{12_listallchannels}. In the
determination of the detection efficiencies, the branching fractions
of the unstable intermediates (eg. $\Lambda$, $\bar{\Sigma}^+$) are
included by generating all their possible decay modes in the
corresponding MC samples.

\begin{figure*}[htbp]
{
       \includegraphics[width=7cm,height=5cm]{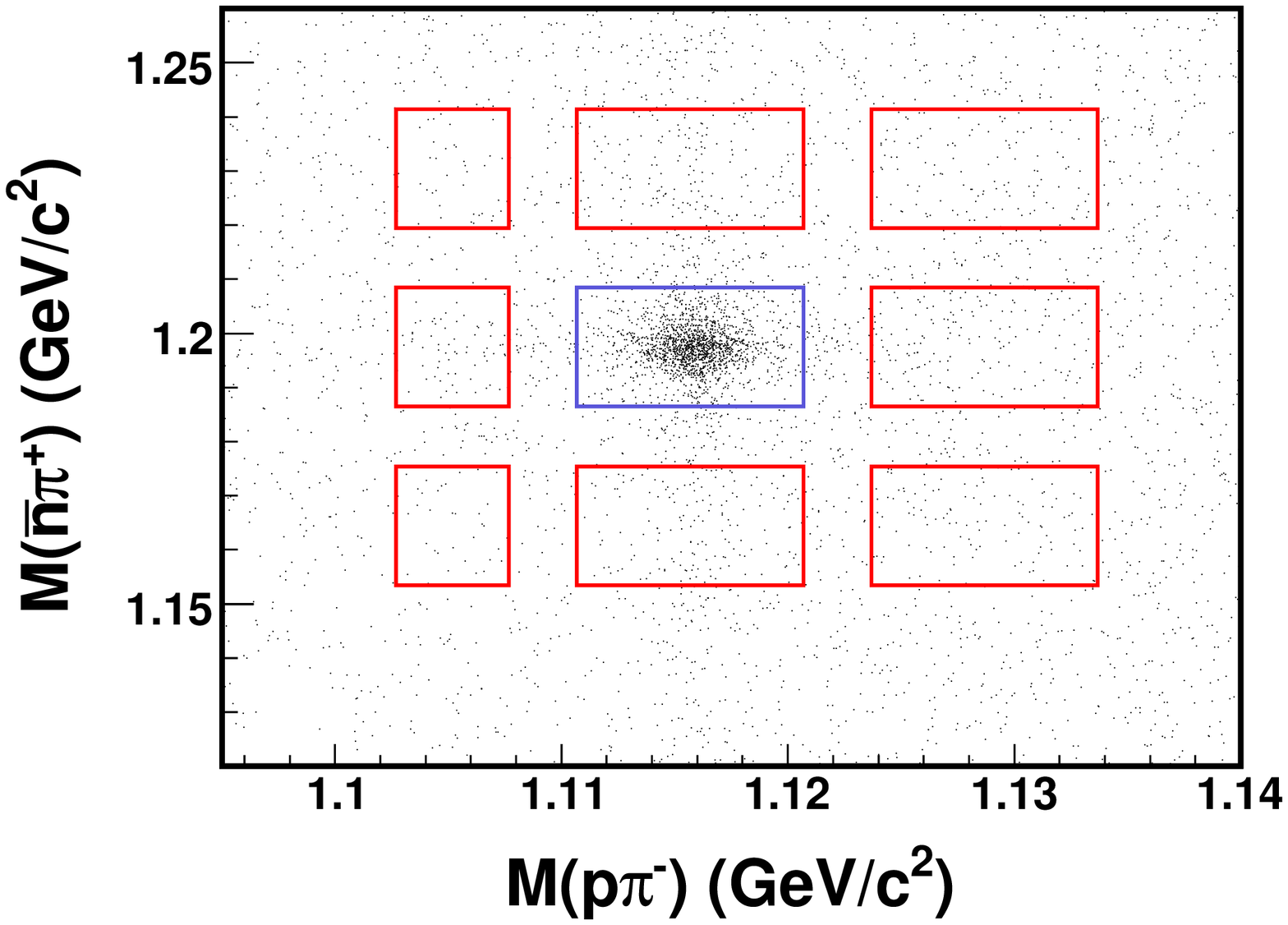}\put(-25,100){(a)}
      \label{1_scatter_LS}
}
{
      \includegraphics[width=6.5cm,height=5cm]{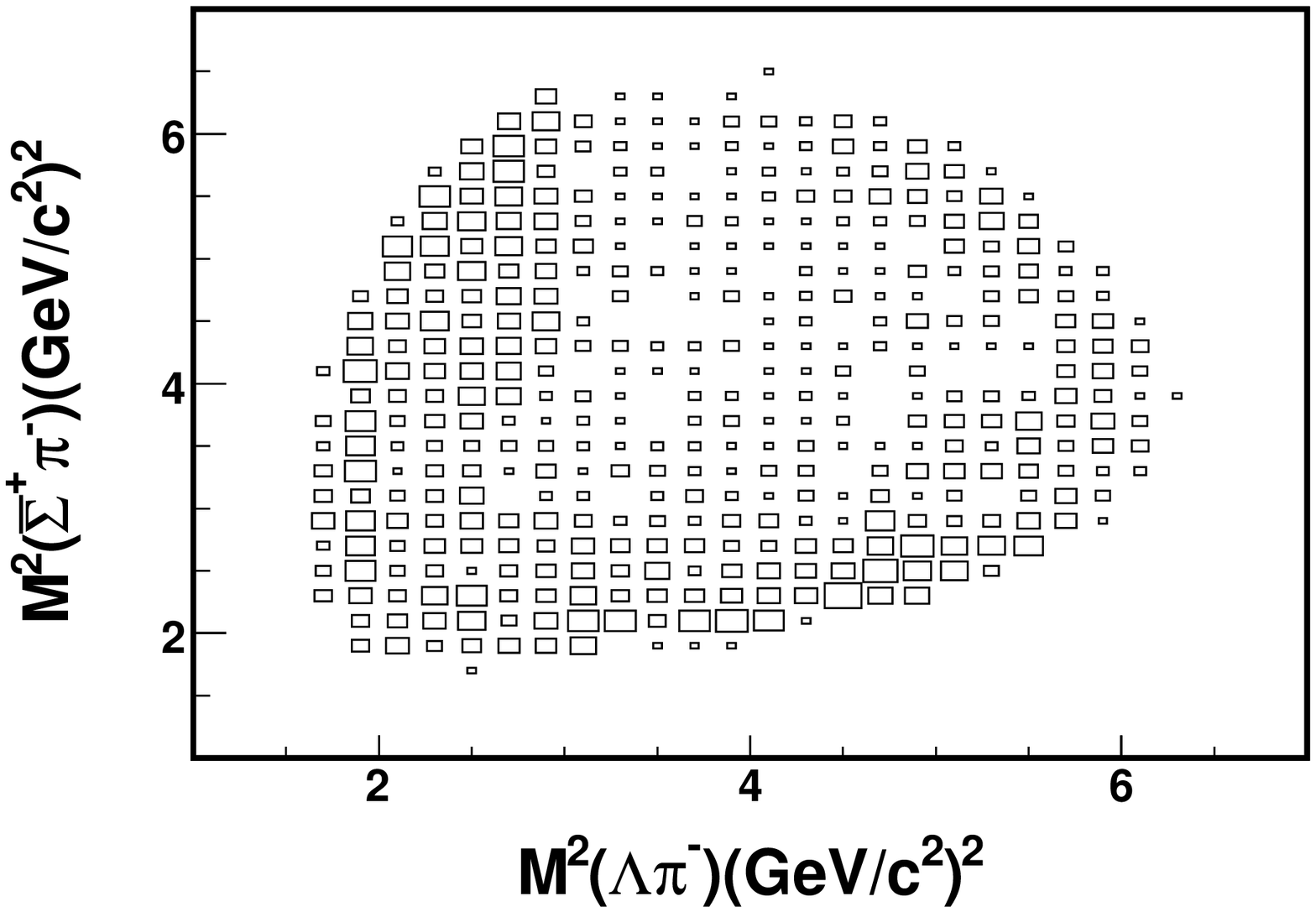}\put(-25,100){(b)}
      \label{1_pwafit_dalitzData}
}
  \caption{(a) The scatter plot of $M(p\pi^-)$ versus
$M(\bar{n}\pi^+)$, where the boxes denote the signal regions and
the sideband regions for background estimation; (b) the Dalitz plot of $\psi(3686)\rightarrow
{\Lambda} \bar\Sigma^+\pi^-$ candidate events.}
\label{1_pwafit_dalitzData}
\label{1_scatter_LS}
\end{figure*}

\begin{figure*}[htbp]
 {
      \includegraphics[width=5.cm,height=4.5cm]{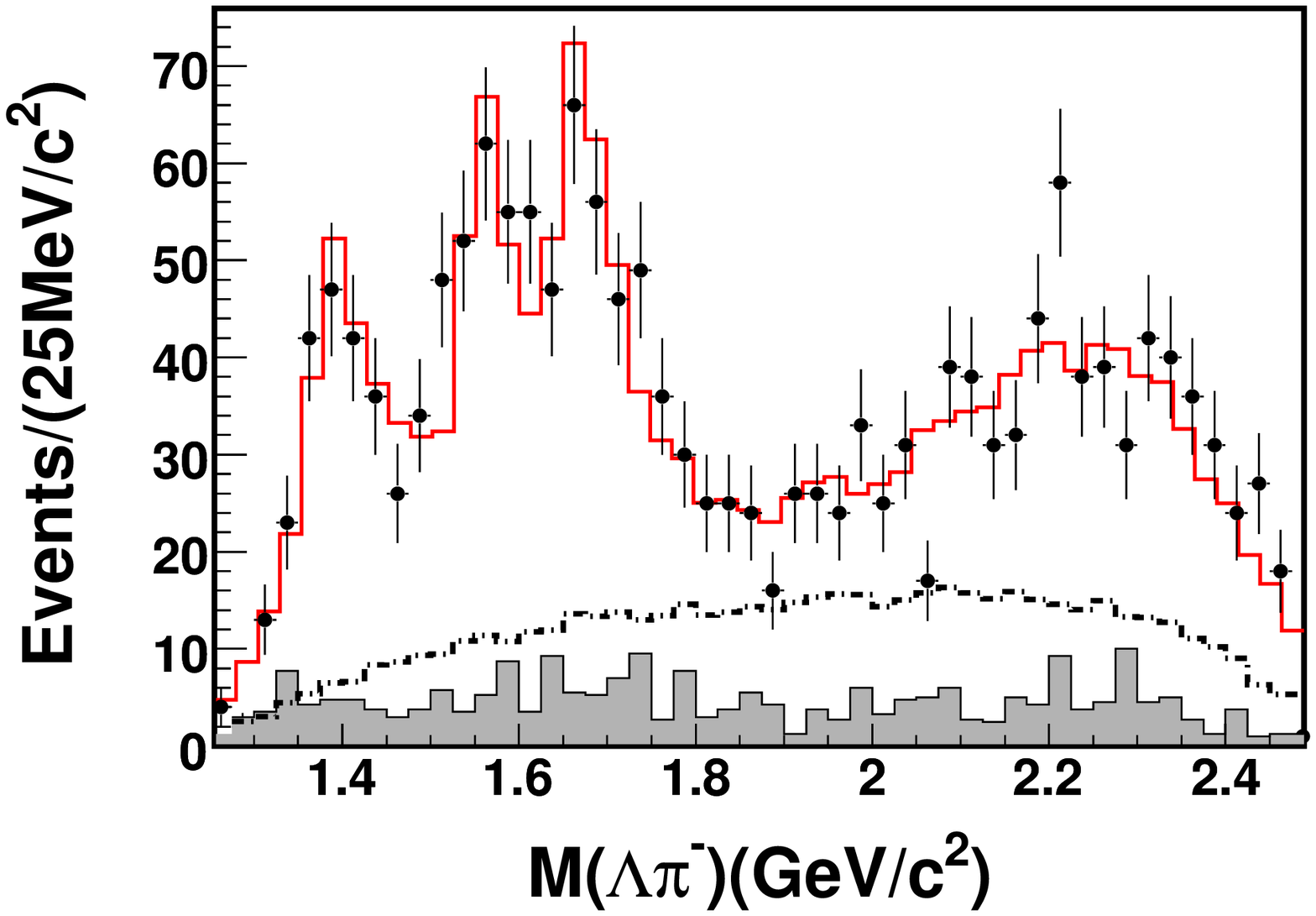}\put(-25,100){(a)}
      \label{1_pwafit_LPi}
} {
      \includegraphics[width=5.cm,height=4.5cm]{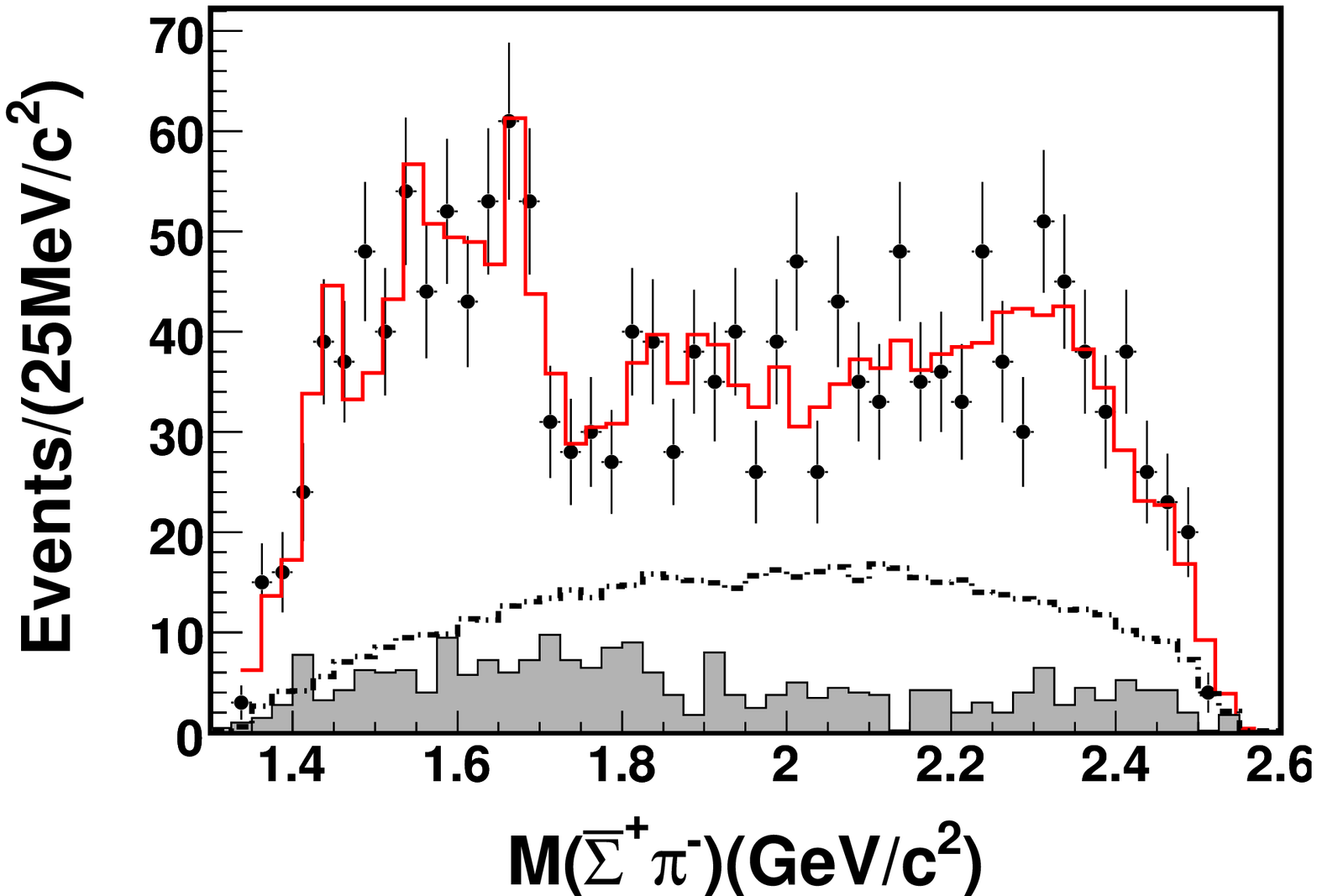}\put(-25,100){(b)}
      \label{1_pwafit_SPi}
}
 {
      \includegraphics[width=5.cm,height=4.5cm]{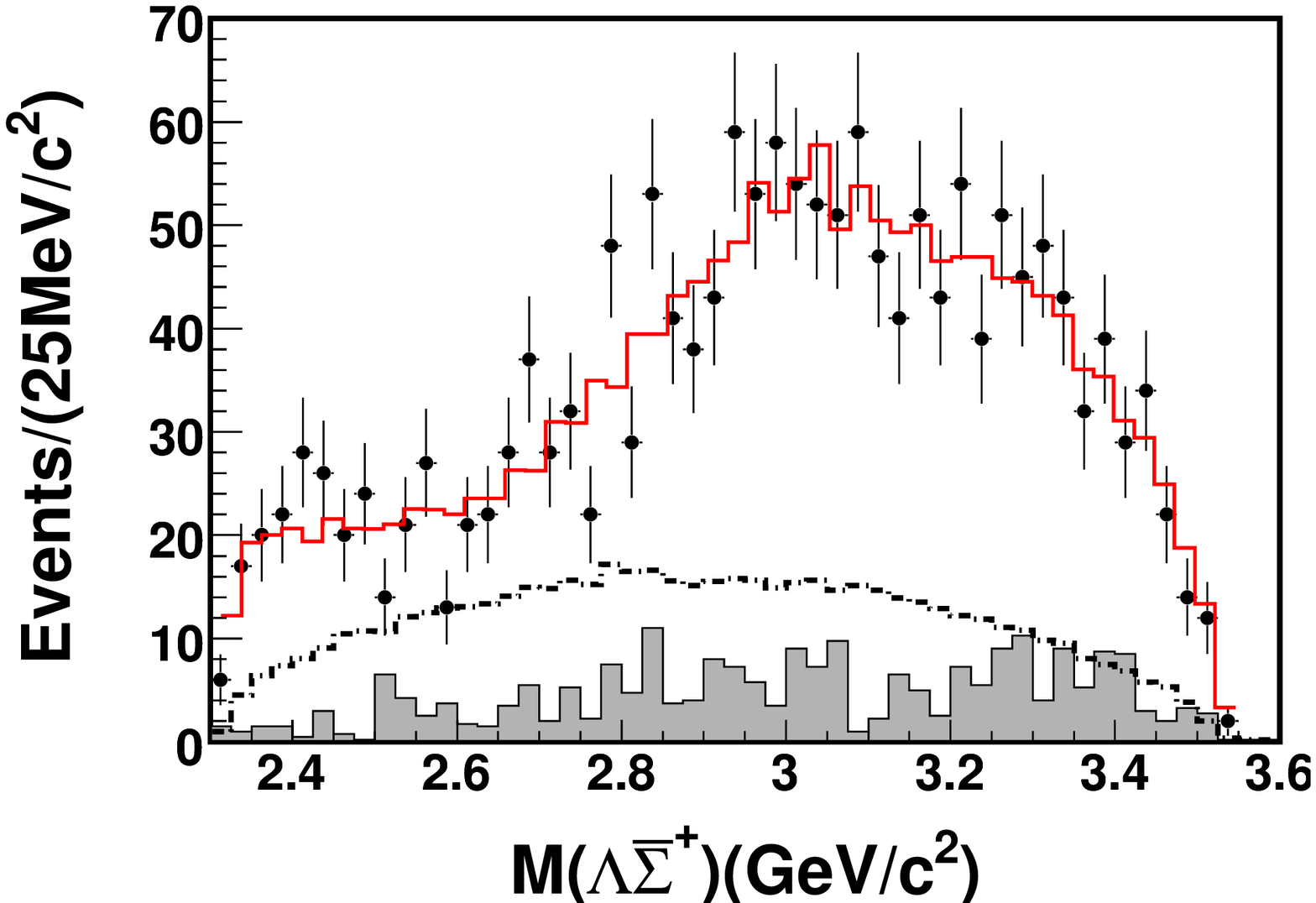}\put(-25,100){(c)}
      \label{1_pwafit_LS}
}
 \caption{Comparisons between data and PWA projections of $\psi(3686)\rightarrow
{\Lambda} \bar\Sigma^+\pi^-$, (a)
   $M(\Lambda\pi^{-})$, (b) $M(\bar\Sigma^{+}\pi^{-})$ and (c)
   $M(\Lambda\bar\Sigma^{+})$.  Points with error bars are
   data, the solid histograms are PWA projections, the dashed
   histograms are phase space distributions from MC simulation, and
   the shaded histograms are the background contributions estimated
   from the $\Lambda-\bar{\Sigma}$ sidebands. }
\label{1_pwafit_LPi}
\label{1_pwafit_SPi}
\label{1_pwafit_LS}
\end{figure*}

\section{Systematic uncertainties}

The systematic uncertainty due to the charged track detection
efficiency has been studied with control samples $J/\psi \rightarrow p
K^- \bar{\Lambda} + c.c.$ and $J/\psi \rightarrow \Lambda
\bar{\Lambda}$ decays. The difference of the charged tracking
efficiencies between data and MC simulation is $2\%$ per track. In
this analysis, there are four charged tracks in the final states, and
the uncertainty is determined to be $8 \%$.

The PID efficiency for MC simulated events agrees with the one
determined using data within $1 \%$ for each proton or anti-proton
according to the study of $J/\psi\rightarrow p\bar
p\pi^{+}\pi^{-}$~\cite{PPPILYT}. $1 \%$ is taken as the uncertainty from PID in each channel. The photon reconstruction efficiency
is studied using the control sample of
$J/\psi\rightarrow\rho^{0}\pi^{0}$ events, as described in
~\cite{PHOTON}. The efficiency difference between data and MC
simulated events is within $1 \%$ for each photon.

In order to estimate the uncertainty due to the fitting range and the
background function in fitting of $\Sigma$, different mass regions
($\bar{\Sigma}^{\pm}\rightarrow \bar{n} \pi^{\pm}$: from [1.12
GeV$/c^{2}$, 1.26 GeV$/c^{2}$] to [1.14 GeV$/c^{2}$, 1.24
GeV$/c^{2}$], $\bar{\Sigma}^- \rightarrow \bar{p} \pi^{0}$ : from
[1.11 GeV$/c^{2}$, 1.27 GeV$/c^{2}$] to [1.13 GeV$/c^{2}$, 1.25
GeV$/c^{2}$]) have been used to perform the fitting and several
polynomials (from 2nd-order polynomial to 3rd-order) have been used to
describe the backgrounds. The changes of the fitting results are
treated as the corresponding systematic errors.

The uncertainty associated with the 4C kinematic fit is estimated to
be $1.7\%$ using the control sample of
$\psi(3686)\rightarrow\pi^{+}\pi^{-}J/\psi$, $J/\psi\rightarrow p \bar
p \pi^{0}$, $\pi^{0}\rightarrow \gamma\gamma$. The uncertainty
associated with the 1C kinematic fit is estimated to be $2.0\%$ using
the control sample $\psi(3686)\rightarrow\pi^{+}\pi^{-}J/\psi$,
$J/\psi\rightarrow p \bar n \pi^{-}$.

For the detection efficiency derived from the PWA, another MC sample
is generated with only six dominant intermediate excited baryon
states($\Lambda(1116)$, $\Lambda(1520)$, $\Lambda(1670)$, $\Sigma(1385)$, $\Sigma(1580)$, $\Sigma(1670)$), and the difference of the detection efficiencies obtained from
the two different MC samples is taken as the uncertainty from
intermediate excited states.

The uncertainties of the branching fractions are $0.78\%$ for $\Lambda
\rightarrow p \pi$, $0.58\%$ for $\Sigma^{+}\rightarrow p \pi^{0}$,
$0.62\%$ for $\Sigma^{+}\rightarrow n \pi^{+}$, $0.01\%$ for
$\Sigma^{-}\rightarrow n\pi^{-}$ and $0.04\%$ for
$\pi^{0}\rightarrow\gamma\gamma$~\cite{PDG}. The number of
$\psi(3686)$ events is determined to be
$106.41\times(1.00\pm0.81\%)\times10^6$ with the inclusive hadronic
events, and its uncertainty is $0.81\%$~\cite{NUMBER}.

The sources of the systematic errors discussed above and the
corresponding contributions in the error on the branching fractions
are summarized in Table~\ref{11_sum_systematic error}.  The total
systematic errors are obtained by adding the contributions from all
sources in quadrature.

\begin{table*}[htbp]
{\caption{ The branching fractions and the values used in the
calculation for each decay mode, where the first error is statistic error and the second is systematic one.} \vskip+0.5cm
\label{12_listallchannels}} \centering
\begin{tabular}{lllllll} \hline
\hline
$\psi(3686)\rightarrow$         &$N_{obs}$     &$N_{sid}$   &$N_{QED}$  &$\varepsilon(\%)$
&$\mathcal{B}(\times 10^{-5})$  \\ \hline
$\Lambda\bar\Sigma^{+}\pi^{-}(\bar\Sigma^{+}\rightarrow\bar{n}\pi^+)$       &$1594\pm48$   &$43\pm10$   &$64\pm16$  &$20.25\pm0.15$   &$6.91\pm0.25 \pm0.65$ \\
$\bar\Lambda\Sigma^{-}\pi^{+}(\Sigma^{-}\rightarrow n\pi^-)$      &$1637\pm47$  &$44\pm10$  &$54\pm14$  &$20.55\pm0.15$  &$7.05\pm0.24 \pm0.61$  \\ \hline
$\Lambda\bar\Sigma^{-}\pi^{+}(\bar\Sigma^{-}\rightarrow\bar{n}\pi^-)$   &$898\pm35$   &$28\pm6$   &$25\pm12$  &$10.03\pm0.11$   &$7.93\pm0.36 \pm0.70$  \\
$\bar\Lambda\Sigma^{+}\pi^{-}(\Sigma^{+}\rightarrow n\pi^+)$     &$891\pm35$   &$29\pm6$   &$32\pm11$  &$10.22\pm0.11$  &$7.64\pm0.35 \pm0.69$  \\ \hline
$\Lambda\bar\Sigma^{-}\pi^{+}(\bar\Sigma^{-}\rightarrow\bar{p}\pi^{0})$    &$458\pm23$   &$18\pm5$   &$26\pm10$  &$5.34\pm0.078$  &$7.29\pm0.47 \pm0.72$  \\
$\bar\Lambda\Sigma^{+}\pi^{-}(\Sigma^{+}\rightarrow p\pi^{0})$    &$554\pm26$   &$13\pm5$   &$33\pm11$  &$6.22\pm0.081$  &$7.68\pm0.67 \pm0.71$  \\
\hline \hline
\end{tabular}
\end{table*}

\section{Results}
For the decays analyzed in this analysis, the branching fractions
are obtained using the following formula:

\begin{equation}
\mathcal{B}(\psi(3686)\rightarrow
\Lambda\bar\Sigma^{+}\pi^{-} (\Lambda \bar\Sigma^-\pi^+) )=
\frac{N_{obs}-N_{sid}-N_{QED}}{N_{\psi(3686)}\times \varepsilon},
\end{equation}
where $N_{obs}$ is the number of observed
$\bar{\Sigma}^+(\bar\Sigma^-) $ events, $N_{sid}$ is the number of
background events estimated from $\Lambda$ sidebands, $N_{QED}$ is the
number of background events from QED processes, $\varepsilon$ is the
detection efficiency obtained from the MC simulation after accounting
for the branching factions of intermediate states, and
$N_{\psi(3686)}$ is the number of $\psi(3686)$ events, which is
determined from the inclusive hadronic events~\cite{NUMBER}.

The resulting branching fractions are summarized in
Table~\ref{12_listallchannels}, in which the first errors are
statistical and the second ones systematic.

\begin{table*}[htbp]
{\caption{Summary of systematic sources and the corresponding contributions (\% ).} \vskip+0.5cm \label{11_sum_systematic error}} \centering
\begin{tabular}{lcccccc} \hline
\hline

$\psi(3686)\rightarrow$  &$\Lambda\bar\Sigma^{+}\pi^{-}$  &$\Lambda\bar\Sigma^{-}\pi^{+}$ &$\Lambda\bar\Sigma^{-}\pi^{+}$
&$\bar\Lambda\Sigma^{+}\pi^{-}$ &$\bar\Lambda\Sigma^{+}\pi^{-}$ &$\bar\Lambda\Sigma^{-}\pi^{+}$  \\
Sources  &$(\bar\Sigma^{+}\rightarrow\bar{n}\pi^+)$    &$(\bar\Sigma^{-}\rightarrow\bar{n}\pi^-)$   &$(\bar\Sigma^{-}\rightarrow\bar{p}\pi^{0})$  &$(\Sigma^{+}\rightarrow n\pi^+)$   &$(\Sigma^{+}\rightarrow p\pi^{0})$   &$(\Sigma^{-}\rightarrow n\pi^-)$  \\ \hline
Track detection efficiency  &8  &8  &8  &8  &8  &8 \\
Particle identification   &1  &1  &1  &1  &1  &1 \\
Photon detection efficiency &--  &--  &2  &--  &2  &--  \\
Fitting of $\Sigma$ mass  &3.6  &2.7  &0.7  &3.1  &2.6  &1.5  \\
Kinematic fit  &2.0  &2.0  &1.7  &2.0  &1.7  &2.0  \\
Intermediate excited states  &2.3  &0.1  &5.1  &1.0  &2.2  &1.4  \\
$\mathcal{B}(\Lambda \rightarrow p \pi^-)$  &0.78  &0.78  &0.78  &0.78  &0.78  &0.78  \\
$\mathcal{B}(\Sigma^+\rightarrow n \pi^+ \ $or$ \ p \pi^0)$  &0.005  &0.62  &0.58  &0.62  &0.58  &0.005  \\
$\pi^{0}\rightarrow\gamma\gamma$  &--  &--  &0.034  &--  &0.034  &--  \\
Number of $\psi(3686)$ events &0.81 &0.81 &0.81 &0.81 &0.81 &0.81  \\ \hline
Total  &9.4  &8.8 &9.9 &9.0  &9.2  &8.6  \\
\hline
\hline
\end{tabular}
\end{table*}

\section{Summary}\label{Summary}

Based on 106 million $\psi(3686)$ events collected with the BESIII
detector, the decays $\psi(3686)\rightarrow \Lambda \bar\Sigma^{+}
\pi^{-} +c.c.$ and $\psi(3686)\rightarrow \Lambda \bar\Sigma^{-}
\pi^{+} +c.c.$ are analyzed, and excited strange baryons (eg. peaks
around $1.5$ GeV$/c^{2}$ to $1.7$ GeV$/c^{2}$ in the invariant mass
spectra of $\bar\Sigma^+\pi^-$ and $\Lambda\pi^-$) are observed. The
branching fractions are measured for the first time and summarized in
Table~\ref{12_listallchannels}.  For each decay mode, the branching
fraction is in good agreement with its charge-conjugate reaction. With
the approach proposed in Ref.~\cite{COM}, the weighted average of the
measurements are determined to be
\begin{center}
$\mathcal{B}(\psi(3686) \rightarrow \Lambda\bar\Sigma^{+}\pi^{-} + c.c.) = (1.40\pm 0.03 \pm 0.13 )\times10^{-4}$,
\\
$\mathcal{B}(\psi(3686) \rightarrow \Lambda\bar\Sigma^{-}\pi^{+} + c.c.) = (1.54\pm 0.04 \pm 0.13 )\times10^{-4}$,
\end{center}
where the first errors are statistical and the second ones systematic.

With the branching fraction of
$J/\psi\rightarrow\Lambda\bar\Sigma^-\pi^+$~\cite{PDG}, we obtain:
\begin{equation}
Q_{\Lambda\bar\Sigma^-\pi^+}=
\frac{\mathcal{B}(\psi(3686)\rightarrow
\Lambda\bar\Sigma^{-}\pi^{+})}{\mathcal{B}(J/\psi\rightarrow
\Lambda\bar\Sigma^{-}\pi^{+})} =  (9.3\pm1.2) \%,
\end{equation}
which tests the ``$12\%$ rule'' for this decay.

\section{Acknowledgments}

  The BESIII collaboration thanks the staff of BEPCII and the computing center for their hard efforts.
  This work is supported in part by the Ministry of Science and Technology of China under Contract No. 2009CB825200;
  National Natural Science Foundation of China (NSFC) under Contracts Nos. 10625524, 10805053, 10821063, 10825524, 10835001, 10935007, 11125525, 10979038, 11079030, 11005109;
  Joint Funds of the National Natural Science Foundation of China under Contracts Nos. 11079008, 11179007, 10979012, U1232107;
  the Chinese Academy of Sciences (CAS) Large-Scale Scientific Facility Program; CAS under Contracts Nos. KJCX2-YW-N29, KJCX2-YW-N45;
  100 Talents Program of CAS; Istituto Nazionale di Fisica Nucleare, Italy; Ministry of Development of Turkey under Contract No. DPT2006K-120470;
  U. S. Department of Energy under Contracts Nos. DE-FG02-04ER41291, DE-FG02-91ER40682, DE-FG02-94ER40823; U.S. National Science Foundation;
  University of Groningen (RuG); the Helmholtzzentrum fuer Schwerionenforschung GmbH (GSI), Darmstadt;
  and WCU Program of National Research Foundation of Korea under Contract No. R32-2008-000-10155-0.


\end{document}